\documentclass[aps,pra,twocolumn,showpacs,superscriptaddress]{revtex4-1}
\usepackage{graphicx}    
\usepackage{dcolumn}    
\usepackage{bm}             
\usepackage{amssymb}   
\usepackage{amsmath}    
\usepackage{subfigure}    
\usepackage{braket}
\usepackage{mathrsfs}
\usepackage{xcolor}
\usepackage[export]{adjustbox}
\usepackage[toc,page]{appendix}
\begin{document}

\title{Analogue quantum simulation of superradiance and subradiance in trapped ions}

\author{R. T. Sutherland}
\email{sutherland11@llnl.gov}
\affiliation{Physics Division, Physical and Life Sciences, Lawrence Livermore National Laboratory, Livermore, California 94550, USA}

\date{\today}

\begin{abstract} 
We discuss a protocol for the analogue quantum simulation of superradiance and subradiance using a linear chain of $N$ trapped qubit ions with a single sympathetic cooling ion. We develop a simple analytic model that shows the dynamics of the qubit subspace converge to those of a cloud undergoing Dicke superradiance and subradiance. We provide numerical simulations of the full ion chain and show that they converge to the dynamics predicted by our analytic model with no fitting parameters. We also map out the parameter regime needed to reach this convergence. 
\end{abstract}
\maketitle

\section{Introduction}
\begin{figure}[b]
\includegraphics[width=0.5\textwidth]{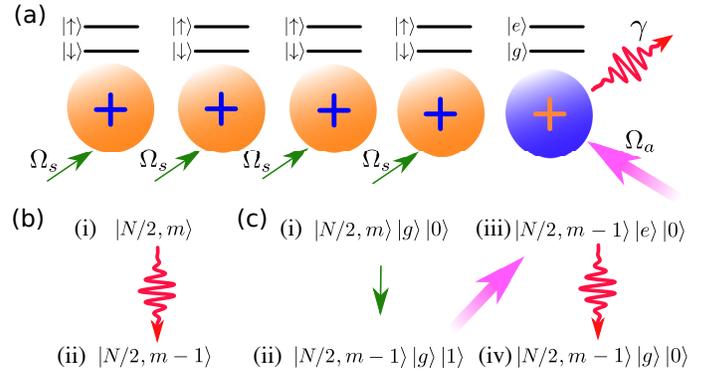}
\centering
\caption{(a) Illustration of the trapped ion system to be used as an analogue quantum simulation of superradiance and subradiance. The system consists of $N$ qubit ions (orange), symmetrically coupled to the motion with a field gradient (green arrows). The motion is also coupled to an auxiliary ion with a laser (pink arrow) and emits radiation (red arrow). (b) Illustration of a photon decay for Dicke superradiance. Here, the Dicke state with $\frac{N}{2} + m$ excitations, $\ket{N/2,m}$, emits a single photon and is projected onto the Dicke state with $\frac{N}{2} + m-1$ excitations, $\ket{N/2,m-1}$. (c) Illustration of the protocol used to mimic superradiance. The system, initially in the $\ket{N/2,m}\ket{g}\ket{0}$ state, is symmetrically driven to the $\ket{N/2,m-1}\ket{g}\ket{1}$ state, which is quickly driven to the $\ket{N/2,m-1}\ket{e}\ket{0}$ state, the system then radiates to the $\ket{N/2,m-1}\ket{g}\ket{0}$ state; the qubit subspace follows the trajectory described in (b).}
\label{fig:fig_1}
\end{figure}

Trapped ions are one of the most promising platforms for quantum simulations \cite{wineland_1998,leibfried_2002,nielsen_2010,cirac_1995, monroe_1995}. Using either lasers \cite{cirac_1995,monroe_1995,ballance_2016,gaebler_2016} or microwaves \cite{mintert_2001,ospelkaus_2008,ospelkaus_2011,harty_2014,harty_2016,wunderlich_2017,hahn_2019,srinivas_2018,sutherland_2019} to generate spin-motion coupling, they offer the ability to create a large range of effective Hamiltonians that have controllable spin-spin interactions \cite{porras_2004,kim_2009} and environmental coupling \cite{barreiro_2011}. This has enabled analogue quantum simulations of important many-body phenomena such as superradiant phase-transitions \cite{safavi-naini_2018} quantum transport \cite{gorman_2018}, and the Dirac equation \cite{gerritsma_2010}. There has yet to be an analogue quantum simulation, however, of superradiance and subradiance \cite{dicke1954}, which are essential parts of many physical processes, including atomic, biological, and condensed matter systems \cite{gross_1982,ido2005,bettles_2016, schneble_2003, sutherland_2016_1,sutherland_2016_2,sutherland_2017_3,sutherland_2017_4,scully_2006, kaiser_2018,bromley_2016,lee_2016,jennewein_2016,pellegrino2014,roof_2016,molina_2016,monshouwer_1997,scheibner_2007,baumann_2010,bradac_2017,delanty_2011,mlynek_2014}. While the study of coherent emission from quantum systems has yielded important results in its tenure, the Hilbert space describing a typical system of interest will grow exponentially with its size. This has prompted physicists to create simplified theoretical frameworks by making physical approximations such as assuming a low \cite{ruostekoski1997,lee_2016,sutherland_2017_4} or high \cite{sutherland_2017_3} fraction of the atoms to be excited, or by leveraging symmetries in interparticle interactions \cite{xu_2013}. In order to explore larger and less constrained systems, a quantum simulation for coherent radiation will likely be needed. In this work, we will show that such a simulation is possible in a trapped ion system, where the photon emission from a single auxiliary sympathetic cooling ion in an $N$ qubit system (see Fig.~\ref{fig:fig_1}a), converges to the exact photon emission of an $N$ atom cloud exhibiting Dicke superradiance \cite{dicke1954}, with an effective single atom decay rate that depends on (controllable) field strengths.

In its original formulation, Dicke superradiance is represented by a cloud of $N$ two-level atoms confined to a space that is small compared to the atoms' transition wavelengh \cite{dicke1954}, ignoring the strong dipole-dipole interactions present in this regime \cite{moi_1983,gross_1982}. If all the two-level atoms are placed in the excited state, the system only decays into symmetric superpositions of every possible state with a given number of excitations. Here, using the parallel between two-level and spin-1/2 systems, we represent these \textit{Dicke states} as $\ket{N/2,m}$. In this notation, $m$ is the spin projection onto the $z$-axis of an $N$ spin-1/2 system with total angular momentum $\frac{N}{2}$. As the system decays, correlations between the atoms build, causing the enhanced decay rate of the system to be $\Gamma^{\prime} (\frac{N}{2} + m)(\frac{N}{2} - m + 1)$, as opposed to the $\Gamma^{\prime}(\frac{N}{2}+m)$ rate expected in the absence of correlations, where $\Gamma^{\prime}$ is the single atom decay rate. The increase in emission rate as the system evolves gives the signature $\propto N^{2}$ photon intensity pulse associated with superradiance. This has been shown to be equivalent to an ensemble of two-level atoms symmetrically coupled to a lossy cavity \cite{bonifacio_1971,delanty_2011,mlynek_2014}. It has also been shown that this lossy cavity effect may be generated by sympathetically cooling a Coulomb mode of a mixed-species Penning trap, where the mechanism may be used to effectively create steady-state superradiance \cite{shankar_2017}. In this work, we will use a similar underlying physical mechanism, to create an analogue quantum simulation of the superradiant cascade effect described in Dicke's original work \cite{dicke1954} in a linear chain of ions with a \textit{single} auxiliary ion  (same or different species) used for sympathetic cooling. We will show that, when the motion is cooled significantly faster than the rate of spin-motion coupling, the temporal dependence of the qubit evolution, as well as the scattered radiation of the auxiliary ion, converges to that expected from Dicke superradiance and subradiance. We expect that the work shown here can be expanded to include effective spin-spin interactions, and eventually be used for simulations that are unfeasible on a classical computer.

\section{Theory}\label{sec:theory}

We consider a system of trapped ions that is comprised of $N$ qubit ions, a single motional mode, and an auxiliary ion for sympathetic cooling (see Fig.~\ref{fig:fig_1}a). The elements of the initial representation of the system are written as:
\begin{equation}\label{eq:state_no_dicke}
    \ket{\psi} = \ket{qubit}\ket{aux}\ket{n},
\end{equation}
where $\ket{qubit}\equiv\ket{\uparrow\uparrow\downarrow...\uparrow}$ represents the internal qubit states, $\ket{aux}$ represents the internal state of the auxiliary ion, and $\ket{n}$ represents a motional mode containing $n$ phonons. Here, $\ket{\uparrow}(\ket{\downarrow}$) is the excited(ground) state of an individual qubit. The auxiliary ion is either in the strongly radiating state, $\ket{e}$, or the non-radiating ground state, $\ket{g}$; this is achieved with a laser-driven cycling transition on an ion that may or may not be the same species as the qubits. We assume a system initially cooled to the motional ground state, $n=0$, where the auxiliary ion is in the $\ket{g}$ state. The qubits are then symmetrically driven on the red sideband with a continuous field, while a separate red sideband drive is applied to the auxiliary ion. The Hamiltonian for this system is:
\begin{eqnarray}\label{eq:hamiltonian_big}
    \hat{H} = \hbar\Omega_{s}\Big\{\hat{S}_{+}\hat{a} + \hat{S}_{-}\hat{a}^{\dagger}\Big\} + \hbar\Omega_{a}\Big\{\hat{\sigma}_{+}\hat{a} + \hat{\sigma}_{-}\hat{a}^{\dagger} \Big\}.
\end{eqnarray}
Note that we are working in the rotating frame with respect to the qubit and motional frequency, and have made the rotating wave approximation. Here, $\hat{a}(\hat{a}^{\dagger})$ are the phonon annihilation(creation) operators, $\hat{S}_{z,+,-} \equiv \sum_{j}^{N}\hat{\sigma}_{z,+,-}^{j}$ represents a collective spin operator on the qubit subspace, where $\hat{\sigma}_{z,+,-}^{j}$ is a Pauli spin operator for the $j^{th}$ qubit, and $\hat{\sigma}_{+}(\hat{\sigma}_{-})$ are Pauli spin operators for the auxiliary ion. The amplitudes in this equation are the Rabi frequencies of the red sidebands acting on the qubits, $\Omega_{s}$, and the red sideband acting on the auxiliary ion, $\Omega_{a}$. 

In order to perform the many-ion calculations in Sec.~\ref{sec:superradiance}, we simplify the equations by noting that when the operator, $\hat{S}_{+,-}$, acts on a superradiant state, given by:
\begin{equation}
    \ket{N/2,m} \equiv \Big\{ \frac{(N-m)!m!}{N!}\Big\}^{1/2}\sum_{qubits}\ket{\uparrow \downarrow\downarrow\uparrow...\uparrow},
\end{equation}
where the sum in the above equation is over all states with $\frac{N}{2} + m$ qubits in the excited state, it only couples to $\ket{N/2,m\pm1}$ with non-zero matrix elements. Note that $\ket{N/2,m}$ is an eigenstate of $\hat{S}^{2}$ and $\hat{S}_{z}$. This allows us to greatly reduce the qubit subspace from the $2^{N}$ original elements to the $N+1$ possible $\ket{N/2,m}$ states. We can then rewrite Eq.~(\ref{eq:hamiltonian_big}) as:
\begin{eqnarray}
        \hat{H} &= & \hbar\Omega_{s}\Big\{\hat{D}_{+}\hat{a} + \hat{D}_{-}\hat{a}^{\dagger}\Big\}  + \hbar\Omega_{a}\Big\{\hat{\sigma}_{+}\hat{a} + \hat{\sigma}_{-}\hat{a}^{\dagger} \Big\}, \nonumber \\
\end{eqnarray}
where we have rewritten the qubit operators as:
\begin{eqnarray}
    \hat{D}_{-} \equiv\ \sum_{m}\{(\frac{N}{2} + m)(\frac{N}{2} - m + 1)\}^{1/2}\ket{N/2,m-1}\bra{N/2,m}, \nonumber \\
\end{eqnarray}
and $\hat{D}_{+} = \hat{D}_{-}^{\dagger}$, thereby greatly reducing the complexity of the above equations. We assume that the radiation from the qubits themselves is negligible. The auxiliary ion, however, is taken to radiate quickly. We account for this using the Lindblad formalism for a single two-level system's decay, where the photon bath has been traced over. The full master equation for the density operator of the system, $\hat{\rho}$, is:
\begin{eqnarray}\label{eq:main}
    \dot{\hat{\rho}} &=& -\frac{i}{\hbar}\Big[\hat{H},\hat{\rho}\Big] + \Gamma\Big\{\hat{\sigma}_{-}\hat{\rho}\hat{\sigma}_{+} - \frac{1}{2}\hat{\sigma}_{+}\hat{\sigma}_{-}\hat{\rho} - \frac{1}{2}\hat{\rho}\hat{\sigma}_{+}\hat{\sigma}_{-} \Big\},
\end{eqnarray}
where $\Gamma$ is the decay rate of the auxiliary ion. Note that the full Eq.~(\ref{eq:main}) is used for \textit{all} the numerical calculations of trapped ions in this work, with no further approximations.

\subsection*{Analytic Model}

The superradiant cascade is where the Dicke state, $\ket{N/2,m}$, emits a photon into a Markovian bath and is projected onto $\ket{N/2,m-1}$ at a rate of $\Gamma^{\prime} (\frac{N}{2} + m)(\frac{N}{2} - m + 1)$ (Fig.~\ref{fig:fig_1}b). We simulate the dynamics of this system using trapped ions with the protocol illustrated in Fig.~\ref{fig:fig_1}c. Initially, the qubit subspace is in the $\ket{N/2,m}$ state, the auxiliary ion is in the $\ket{g}$ state, and the motional mode is in the $\ket{0}$ state, making the overal state $\ket{N/2,m}\ket{g}\ket{0}$. The red sideband of each of the qubits is then driven with a rate $\Omega_{s}$ so that the system evolves into the $\ket{N/2,m-1}\ket{g}\ket{1}$ state. Simultaneously, the red sideband of the $\ket{g}\leftrightarrow\ket{e}$ transition is driven at $\Omega_{a}$, a much faster rate than $\Omega_{s}$, taking the system from $\ket{N/2,m-1}\ket{g}\ket{1}$ to $\ket{N/2,m-1}\ket{e}\ket{0}$. Finally, at a rate $\Gamma$ much faster than $\Omega_{s}$, the auxiliary ion decays, taking the state from $\ket{N/2,m-1}\ket{e}\ket{0}$ to $\ket{N/2,m-1}\ket{g}\ket{0}$, repeating this cycle until $\ket{N/2,-N/2}\ket{g}\ket{0}$ is reached. Note that we are assuming the transition $\ket{N/2,m-1}\ket{g}\ket{1}\leftrightarrow \ket{N/2,m-2}\ket{g}\ket{2}$ can be ignored because we are exploring a regime where sympathetic cooling is much faster than this process.

Reference~\cite{dicke1954} derived the rate that population in $\ket{N/2,m}$ decays into $\ket{N/2,m-1}$ (Fig.~\ref{fig:fig_1}b). Similarly, we wish to determine the \textit{effective} decay rate, $\Gamma_{\mathrm{eff}}$, representing the transfer of population from $\ket{N/2,m}\ket{g}\ket{0}$ to $\ket{N/2,m-1}\ket{g}\ket{0}$ (Fig.~\ref{fig:fig_1}c). To do this, we assume an initial population in $\ket{N/2,m}\ket{g}\ket{0}$, and calculate the rate of decay out of the entire $\{\ket{N/2,m}\ket{g}\ket{0}$,$\ket{N/2,m-1}\ket{g}\ket{1}$,$\ket{N/2,m-1}\ket{e}\ket{0}\}$ subspace. We represent the density matrix for this subspace as $\hat{\rho}_{\mathrm{sub}}$. 

To begin, we note that $\hat{H}$ only couples terms \textit{within} $\hat{\rho}_{\mathrm{sub}}$, resulting in no entanglement with the rest of the system. We can therefore focus on the decay of an initial population out of $\hat{\rho}_{\mathrm{sub}}$ in order to determine $\Gamma_{\mathrm{eff}}$. The $\Gamma\hat{\sigma}_{-}\hat{\rho}\hat{\sigma}_{+}$ term in Eq.~(\ref{eq:main}) represents the transfer of population \textit{into} $\hat{\rho}_{\mathrm{sub}}$, which is not a component of the process that we currently wish to analyze; we can thus ignore this term for our analytic model. These considerations allow us to write Eq.~(\ref{eq:main}) for the subspace as:
\begin{equation}\label{eq:reduced_main}
\dot{\hat{\rho}}_{\mathrm{sub}} = -\frac{i}{\hbar}(\hat{H}_{c}\hat{\rho}_{\mathrm{sub}} - \hat{\rho}_{\mathrm{sub}}\hat{H}_{c}^{\dagger}),   
\end{equation}
\noindent where $\hat{H}_{c} \equiv \hat{H} -i\frac{\hbar\Gamma}{2}\hat{\sigma}_{+}\hat{\sigma}_{-}$ is a non-Hermitian Hamiltonian acting on $\hat{\rho}_{\mathrm{sub}}$ \cite{moiseyev_2011}. The above equation is equivalent to a density matrix for a wave function being propagated by $\hat{H}_{c}$. We use this fact to simplify the analytic calculation of $\Gamma_{\mathrm{eff}}$, considering only a \textit{wave function} for the subspace acting under $\hat{H}_{c}$. 

Letting $c_{0}$, $c_{1}$, and $c_{2}$ be the probability amplitudes of the $\ket{N/2,m}\ket{g}\ket{0}$, $\ket{N/2,m-1}\ket{g}\ket{1}$ and $\ket{N/2,m-1}\ket{e}\ket{0}$ states, respectively, we can find the equation of motion for $c_{1}$:
\begin{eqnarray}\label{eq:model_eq_mot}
\dot{c}_{1} &= & - i\Omega_{s}\Big\{\Big(\frac{N}{2} + m\Big)\Big(\frac{N}{2} - m + 1\Big)\Big\}^{1/2}c_{0} - i\Omega_{a}c_{2}. \nonumber \\
\end{eqnarray}
We now assume that (upon reaching quasi-equilibrium) $\dot{c}_{1} \simeq 0$ relative to $\Omega_{a}$ and $\Gamma$. This allows us to set the left hand side of Eq.~(\ref{eq:model_eq_mot}) to $0$ and solve for $c_{2}$ in terms of $c_{0}$:
\begin{eqnarray}\label{eq:c2_solve}
    c_{2} = -\frac{\Omega_{s}\{(\frac{N}{2} + m)(\frac{N}{2} - m + 1)\}^{1/2}c_{0}}{\Omega_{a}},
\end{eqnarray}
where we can now easily solve for $\Gamma_{\mathrm{eff}}$, defined by $\Gamma_{\mathrm{\mathrm{eff}}}(\frac{N}{2} + m)(\frac{N}{2} - m + 1)|c_{0}|^{2} = \Gamma|c_{2}|^{2}$. This results in a final equation:
\begin{eqnarray}\label{eq:gamma_eff}
    \Gamma_{\mathrm{eff}} = \Big(\frac{\Omega_{s}}{\Omega_{a}}\Big)^{2}\Gamma.
\end{eqnarray}
\noindent
This is analogous to an $N$ atom system undergoing superradiance with an effective single atom decay rate of $\Gamma^{\prime} = \Gamma_{\mathrm{eff}}$. 

\section{Results}
\subsection{Superradiance}\label{sec:superradiance}
\begin{figure}[h]
\includegraphics[width=0.45\textwidth]{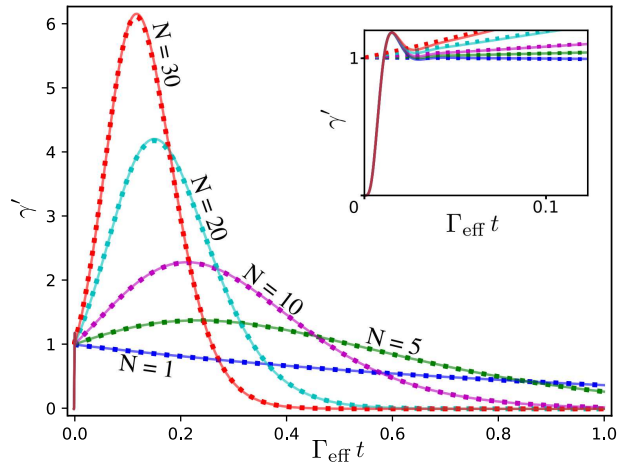}
\centering
\caption{Comparison between simulations of the original $N$ particle system exhibiting superradiance \cite{dicke1954}, and the full simulation (Eq.~(\ref{eq:main})) of the analogue quantum simulator described in this work. For the calculation that uses the original equations (dotted lines), we show the photon emission rate, $\gamma$, of the entire system normalized by the single atom decay rate, $\Gamma_{\mathrm{eff}}$, and the number, $N$, of particles $\gamma^{\prime} \equiv \gamma/N\Gamma_{\mathrm{eff}}$. For the calculation of our trapped ion system (solid lines), we show $\gamma^{\prime}$ from the auxiliary ion for an $N$ qubit system. This shows that after an initial ramp up time (see inset), our protocol converges Dicke superradiance and our analytic model gives an exact prediction of $\Gamma_{\mathrm{eff}}$.}
\label{fig:fig_2}
\end{figure}

For an ensemble exhibiting the form of superradiance described in Ref.~\cite{dicke1954}, a Dicke state, $\ket{N/2,m}$, decays at a rate given by $\Gamma^{\prime} (\frac{N}{2} + m)(\frac{N}{2} - m + 1)$. This indicates that if an ensemble is initialized to the $\ket{N/2,N/2}$ state then the photon emission rate of the ensemble, $\gamma = \Gamma \langle\hat{\sigma}_{+}\hat{\sigma}_{-}\rangle = \mathrm{Tr}\{\Gamma\hat{\sigma}_{+}\hat{\sigma}_{-}\hat{\rho}\}$, will initially \textit{increase} with time, followed by a rapid decrease, when the system runs out of energy, emitting a pulse of photons with intensity $\propto N^2$. In this subsection, we demonstrate that $\gamma$ from a single auxiliary ion in an $N$ qubit ion chain converges to the exact time dynamics expected in a superradiant system with a single atom decay rate of $\Gamma^{\prime} = \Gamma_{\mathrm{eff}}$, when $\Gamma \gg \Omega_{s}$ and $\Omega_{a}^{2}/\Gamma \gg \Omega_{s}$.

The convergence of our trapped ion system to Dicke superradiance is shown in Fig.~\ref{fig:fig_2} for systems of $N = 1,5,10,20$ and $30$. This is done using Eq.~(\ref{eq:main}) without approximation. We here show the temporal evolution of $\gamma$ normalized by $\Gamma^{\prime}$ and $N$, $\gamma^{\prime}\equiv \gamma/N\Gamma^{\prime}$. We compare $\gamma^{\prime}$ for the entire superradiant cloud with $\gamma^{\prime}$ from the auxiliary ion in an $N$ qubit trapped ion system, and set $\Gamma^{\prime}$ to $\Gamma_{\mathrm{eff}}$ for comparison. For the trapped ion system, we choose values of $\Omega_{a} \simeq 81.2\Omega_{s}$ and $\Gamma = 200\Omega_{s}$, chosen to be well-within the regime of convergence (discussed below). At very short times, $\gamma^{\prime}$ from the auxiliary ion in the trapped ion system is close to zero. However, at a timescale $\propto 1/\Gamma$, the system reaches quasi-equilibrium, and, as a result, follows the dynamics discussed in Sec.~\ref{sec:theory}. This initial ``ramp-up" time, is shown in the inset of Fig.~\ref{fig:fig_2}. Here, after a time, $t$, that is short compared to $\Omega_{s}$, each system of $N$ qubits converges to the system they are supposed to simulate. 

\begin{figure}[h]
\includegraphics[width=0.45\textwidth]{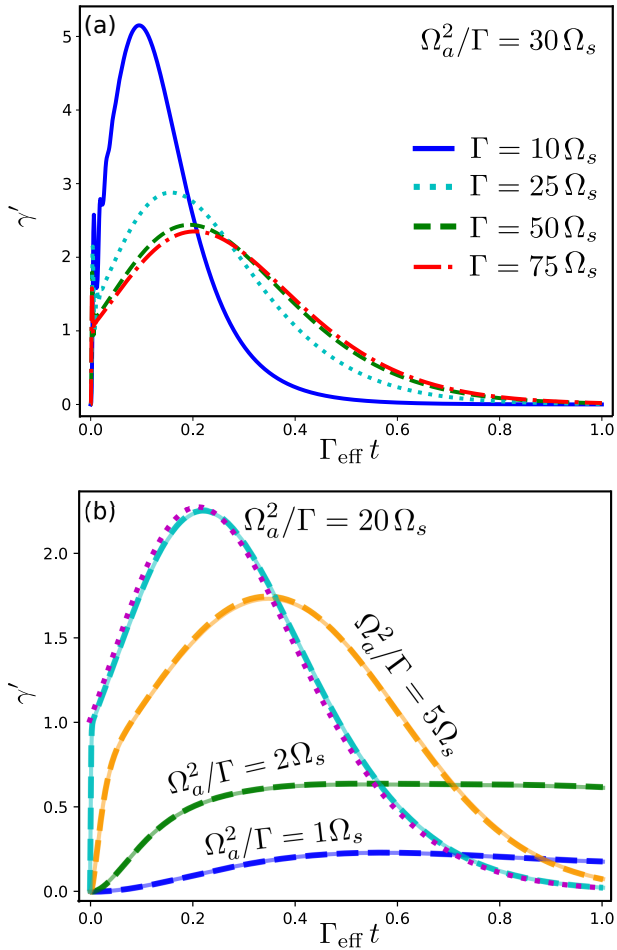}
\centering
\caption{Calculations for $N=10$ qubit systems. (a) Time evolution of the photon emission rate from the auxiliary ion, $\gamma$, normalized by the effective single atom decay rate, $\Gamma_{\mathrm{eff}}$, and the number of qubits, $N$, i.e. $\gamma^{\prime} = \gamma/N\Gamma_{\mathrm{eff}}$. The calculations shown here are for $\Omega_{a}^{2}/\Gamma = 30\Omega_{s}$, and for varying values of $\Gamma$, converging to Dicke superradiance when $\Gamma \gtrsim 75\Omega_{s}$. Note that, while the $\Gamma = 10 \Omega_{s}$ calculation appears to radiate faster than those for larger values of $\Gamma$, it actually radiates slower; this is an artifact of normalizing by $\Gamma_{\mathrm{eff}}$, whose derivation is no longer valid for the calculation. (b) For both auxiliary atom decay rates of $\Gamma = 400\Omega_{s}$ (solid) and $\Gamma = 800\Omega_{s}$ (thick dashed), the emission rate versus $t$ falls on the same line for the same values of $\Omega_{a}^{2}/\Gamma$, converging to Dicke superradiance (purple dotted) for $\Omega_{a}^{2}/\Gamma \gtrsim 20\Omega_{a}$. This shows that when $\Gamma$ is large the dynamics of the system is determined entirely by $\Omega_{a}^{2}/\Gamma$ and $N$.}
\label{fig:fig_3}
\end{figure}

In order for $\gamma^{\prime}$ versus $t$ from the auxiliary ion to converge to that of Dicke superradiance, the parameters of the system must be set up such that the system reaches quasi-equilibrium on a time-scale that is fast compared to the system's overall evolution ($\propto 1/\Gamma_{\mathrm{eff}}$), and that, at quasi-equilibrium, population is removed from the motional mode at a rate that is fast compared to $\Omega_{s}$ such that there is no back-action into the initial state. As we show in Fig.~\ref{fig:fig_3}, these two conditions are met when $\Gamma \gg \Omega_{s}$ and $\Omega_{a}^{2}/\Gamma \gg \Omega_{s}$, respectively. 

The first condition, $\Gamma \gg \Omega_{s}$, is due to the fact that the damping provided by the photon emission from the auxiliary ion will cause the system to reach quasi-equilibrium at a rate $\propto \Gamma$. Since the auxiliary ion will mimic superradiant emission when it has reached quasi-equilibrium (see Sec.~\ref{sec:theory}), it must reach quasi-equilibrium on a timescale that is much shorter than the overall evolution of the system. Also, if the system is not damped at a much faster rate than population is put in, back-action will lead to Rabi oscillations; this will significantly complicate the dynamics.  We show this in Fig.~\ref{fig:fig_3}a, a system of $N=10$ qubits coupled to an auxiliary ion that decays at a rate $\Gamma$. Note that we choose a set value of $\Omega_{a}^{2}/\Gamma = 30\Omega_{s}$ in this graph, so that\textemdash as will be discussed below\textemdash population is removed from the motional mode at a rate that is large enough to reach convergence (see Fig.~\ref{fig:fig_3}b). Here it can be seen that for values of $\Gamma \gtrsim 75 \Omega_{s}$ the evolution converges to that expected from Dicke superradiance, but for smaller values of $\Gamma$ the system deviates from the desired evolution.

The second condition, $\Omega_{a}^{2}/\Gamma \gg \Omega_{s}$, is that once the system has reached quasi-equilibrium, population must be removed from the motional mode ($\ket{N/2,m-1}\ket{g}\ket{1}$ state) at a rate that is significantly faster than it is put in. We can quantify this rate by using a similar argument as was used to obtain Eq.~(\ref{eq:gamma_eff}), but for the reduced Hilbert space of just $\ket{N/2,m-1}\ket{g}\ket{1}$ and $\ket{N/2,m-1}\ket{e}\ket{0}$. As we originally assumed that the population of $\ket{N/2,m}\ket{g}\ket{0}$ was fixed in Eq.~(\ref{eq:model_eq_mot}), we now do the same for $\ket{N/2,m-1}\ket{g}\ket{1}$. This gives:
\begin{equation}
    \dot{c}_{2} = -i\Omega_{a}c_{1} - \frac{\Gamma}{2}c_{2}.
\end{equation}
Setting the left hand side to $0$, we get an effective decay rate out of $c_{1}$ of $4\Omega_{a}^{2}/\Gamma$; this means that, at quasi-equilibrium, the value of $\Omega_{a}^{2}/\Gamma$ must be large compared to $\Omega_{s}$, so that the transfer of population from the motional mode to the quickly radiating state is fast enough for the former to act as a lossy cavity. This is shown in Fig.~\ref{fig:fig_3}b, where we can see that our calculations converge to Dicke superradiance only for values of $\Omega_{a}^{2}/\Gamma \gtrsim 20\Omega_{s}$; for larger values of $\Omega_{a}^{2}/\Gamma$, except for Rabi oscillations at small $t$, we find that the system converges. Figure~\ref{fig:fig_3} also shows that when $\Gamma \gg \Omega_{s}$, the dynamics of the system is entirely dictated by the value of $\Omega_{a}^{2}/\Gamma$. This is here seen in the fact that calculations for systems such that $\Gamma = 200\Omega_{s}$ and $\Gamma = 400\Omega_{s}$ fall on the same lines, reaching convergence when $\Omega_{a}^{2}/\Gamma \gtrsim 20\Omega_{s}$.

In terms of measuring the properties of our superradiant system, one could directly measure the photon emission from the auxiliary ion; for illustrative purposes, we have shown this for the figures in this section. Depending on the particular experiment, however, it may be easier to directly measure the internal states of the qubits using standard techniques \cite{wineland_1998}. This will be particularly useful in when measuring systems that do not radiate strongly, as discussed in the next subsection.

\subsection{Subradiance}\label{sec:subradiance}

\begin{figure}[h]
\includegraphics[width=0.45\textwidth]{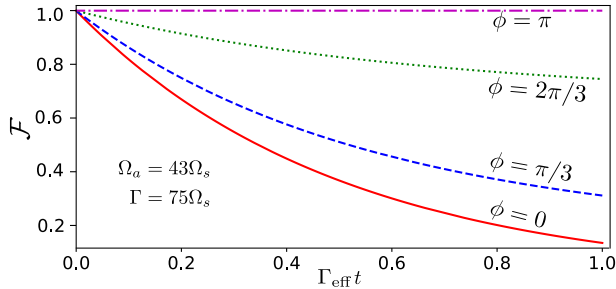}
\centering
\caption{Fidelity, $\mathcal{F}$, versus time, $t$, of a trapped ion system where the $N=2$ qubits are initialized to a state, $2^{-1/2}(\ket{\uparrow\downarrow} + e^{i\phi}\ket{\downarrow\uparrow})\ket{g}\ket{0}$. This illustrates how the state the system is initialized to determines its decay; the ion chain can act as simulator of a superradiant, $\phi = 0$, and subradiant $\phi = \pi$ system.}
\label{fig:fig_4}
\end{figure}

The trapped ion system we describe in this work can mimic the dynamics of subradiance as well. Since subradiance is a significantly harder effect to probe experimentally (due to its diminished emission) \cite{guerin_2016}, this simulation offers a unique opportunity to probe subradiant effects; due to the exceptional readout capabilities of trapped ions \cite{wineland_1998}, one could generate a system that behaves in a subradiant manner, and probe it by simply measuring the internal states of the qubits; this is likely the optimal way to probe this simulation due to the fact that $\gamma^{\prime}$ is small and thus hard to measure when the system mimics subradiance.

In order to demonstrate the capacity of our system to simulate subradiance, Fig.~\ref{fig:fig_4} shows the fidelity, $\mathcal{F} \equiv \bra{\psi_{0}}\hat{\rho}\ket{\psi_{0}}$, versus $\Gamma_{\mathrm{eff}}t$ for a $2$ qubit system placed in the initial state: 
\begin{equation}
    \ket{\psi_{0}} \equiv \frac{1}{\sqrt{2}}\Big\{\ket{\uparrow\downarrow} + e^{i\phi}\ket{\downarrow\uparrow} \Big\}\ket{g}\ket{0}.
\end{equation}
This can be generated using a two-qubit entangling gate to create a Bell state followed by a single qubit rotation \cite{harty_2014,ballance_2016,gaebler_2016,harty_2016}. Note that, here, we simulate the system in the original qubit basis, see Eq.~(\ref{eq:hamiltonian_big}), since $\ket{\psi_{0}}$ is outside of the described $\ket{N/2,m}$ basis. Here we change the value of $\phi$ to vary the state from superradiant ($\phi = 0$), where the state decays at twice the single particle decay rate ($\mathcal{F}(t) = \exp\{-2\Gamma_{\mathrm{eff}}t\}$) to subradiant ($\phi = \pi$), where the initial state does not decay at all $\mathcal{F}(t) = 1$). Even though, in the (idealized) latter case, the auxiliary ion does not emit any radiation, this effect should be observable through the direct measurement of the qubits.

\section{Experimental implementations}
The two parameter regimes required to generate Dicke superradiance ($\Gamma \gtrsim 75\Omega_{s}$ and $\Omega_{a}^{2}/\Gamma \gtrsim 20\Omega_{s}$) can, in principle, be met for any choice of $\Gamma$ and $\Omega_{a}$ by making $\Omega_{s}$ sufficiently weak. This could, of course, result in a small $\Gamma_{\mathrm{eff}}$ which would lead to a long experimental run time, so it is important to determine what this time would be in a typical experiment. Assuming that $\Omega_{a}^{2}/\Gamma \simeq 20\Omega_{s}$, we find $\Gamma_{\mathrm{eff}} \simeq \Omega_{a}^{2}/400\Gamma$, and that this condition automatically enforces $\Gamma \gtrsim 75\Omega_{s}$ for typical values of $\Gamma$. As an example, if the auxiliary ion is $^{9}\mathrm{Be}^{+}$ the value of $\Gamma$ would be approximately $2\pi\times 20~\mathrm{MHz}$ \cite{poulsen_1975}, corresponding to the $P_{3/2}\rightarrow S_{1/2}$ transition. For a value of $\Omega_{a} = 2\pi\times 500~ \mathrm{kHz}$, this would make the convergence criteria for $\Omega_{s} \simeq 2\pi \times 625 ~\mathrm{Hz}$ and the timescale of the experiment $1/\Gamma_{\mathrm{eff}} \simeq 5 ~\mathrm{ms}$. The speed of this scheme is limited by the fact that a typical sympathetic cooling transition decays quickly and requires a strong laser to sufficiently drive the red sideband. The timescale of a potential experiment as well as the required $\Omega_{a}$ could be decreased if one had access to a cooling transition with a \textit{smaller} value of $\Gamma$. 

\section{Conclusion}

In this work, we showed that a trapped ion chain can be made to exhibit Dicke superradiance and subradiance, in a manner that converges to the exact results from the original system. This offers the opportunity to study the phenomena in a way that allows experimentalists to control the effective single atom decay rate, $\Gamma_{\mathrm{eff}}$, by adjusting the field amplitudes of the system. One could even stop the decay of the system at any time and probe any of its particles. For illustrative purposes, we have focused on the simulation of the system originally envisioned by Dicke in Ref.~\cite{dicke1954}, since its symmetries allow us to compare our analogue to the exact answer. Going forward, these ideas could be combined with existing techniques that have used trapped ions to create analogue quantum simulations of other interesting systems \cite{leibfried_2002,gerritsma_2010,maier_2019,safavi-naini_2018,kim_2009,porras_2004,gorman_2018}.  Finally, with various theoretical and technical advances, it could be used to explore the physics of systems not easily probed experimentally or simulated on a classical computer.
\\

\section{Acknowledgements}
The author would like to thank F. Robicheaux, D.H. Slichter, R. Srinivas. M.T. Eiles, S.B. Libby, J. L. DuBois, and S.C. Burd for helpful discussions. Special thanks to F. D. Malone, S.C. Burd, and Y. Rosen for careful reading of the manuscript. Part of this work was performed under the auspices of the US Department of Energy by Lawrence Livermore National Laboratory under Contract DE-AC52-07NA27344. LLNL-JRNL-791496-DRAFT

\section*{References}
\bibliography{biblio}

\end{document}